\documentclass[12pt]{amsart}
\usepackage{amssymb}
\usepackage{amsmath,amsthm,amsfonts}
\usepackage[applemac]{inputenc}
\usepackage{graphicx}
\usepackage{bbm}
\usepackage{pdfsync}
\usepackage{fancyhdr}
\usepackage{mathrsfs}
\usepackage{slashed}
\usepackage{color}
\usepackage{xcolor}
\usepackage[breaklinks,colorlinks,backref]{hyperref}

\hypersetup{
colorlinks, 
linktoc=all, 
linkcolor=blue, 
citecolor=hyptxt,
urlcolor=blue}
\hypersetup{
citebordercolor=,
filebordercolor=green,
linkbordercolor=blue
}
\definecolor{hyptxt}{rgb}{0.7, 0.4, 0.9}

\usepackage{bibtopic}

\usepackage[full]{textcomp} 
\usepackage{fbb} 
\usepackage[varqu,varl]{inconsolata}
\usepackage[libertine,bigdelims,vvarbb]{newtxmath}
\usepackage[cal=boondoxo]{mathalfa}
\usepackage[T1]{fontenc}
\useosf

\newtheorem{prop}{Proposition}[section]

\newcommand{\beprop}{\begin{prop}}
\newcommand{\enprop}{\end{prop}}
\newcommand{\bprf}{\begin{proof}}
\newcommand{\eprf}{\end{proof}}

\newcommand{\ket}[1]{|\kern.3ex#1\kern.3ex\rangle}
\newcommand{\bra}[1]{\langle\kern.3ex #1 \kern.3ex|}
\newcommand{\scalar}[2]{\langle\kern.3ex #1 \kern.3ex|\kern.3ex#2\kern.3ex\rangle}

\def\ii{\mathrm{i}}
\def\ud{\mathrm{d}}

\def\ud{\mathrm{d}}
\definecolor{hervecolor}{rgb}{0.8,0,0.7}

\numberwithin{equation}{section}

\def\1{\mbox{I\hspace{-.15em}1}}

\def\b{\begin{equation}}
\def\e{\end{equation}}

\begin{document}
\date{\today}
\title{``Krein'' regularization method}
\author{M.V. Takook}
\address{\emph{ APC, UMR $7164$}\\
\emph{Universit\'e de Paris} \\
\emph{$75205$ Paris, France}}
\email{ takook@apc.in2p3.fr}
\maketitle
\begin{abstract}
The ``Krein'' regularization method of quantum field theory is studied, inspired by the Krein space quantization and quantum metric fluctuations. It was previously considered in the one-loop approximation, and this paper is generalized to all orders of perturbation theory. We directly recover the physical results previously obtained starting from the standard QFT by imposing the renormalization conditions. By applying our approach to the QFT in curved space-time and quantum linear gravity, we discuss that there is no need for the higher derivative of the metric tensor for the renormalization of the theory. The advantage of our method compared to the previous ones is that the linear quantum gravity is renormalizable in all orders of perturbation theory.
\end{abstract}
\vspace{0.5cm}
{\it Proposed PACS numbers}: 04.62.+v, 98.80.Cq, 12.10.Dm

\tableofcontents

\section{Introduction}

The appearance of singularities in QFT manifests the existence of anomalies in this theory. The normal ordering procedure and the regularization and renormalization techniques are applied to QFT for eliminating these anomalies and obtaining finite results. Although these techniques are efficient in eliminating the singularities and explaining the experimental results with high accuracy, they do not belong to the framework of quantum theory. Moreover, these techniques cannot be applied to quantum field theory in curved space-time and quantum linear gravity either, regardless of curvature, {\it i.e.} they are not renormalizable.

Negative norm states (negative energy solutions of the field equation) were first considered by Dirac in $1942$ to deal with these anomalies \cite{dirac}: {\it ``The appearance of divergent integrals with odd $n$-values in Heisenberg and Pauli's form of quantum electrodynamics may be ascribed to the unsymmetrical treatment of positive- and negative-energy photon states.''} Then, he presented an interpretation for the negative and positive probability amplitude. Muto et Inoue in $1950$ showed that Dirac's proposal of indefinite metric quantization has failed to eliminate all divergences in the hole theory. Moreover, there is difficulty concerning the physical interpretation of the negative probability \cite{muin}.

In $1950$, Gupta applied the idea of indefinite metric quantization to the QED for obtaining a covariant formalism \cite{gupta}. He quantized the radiation field by introducing four types of photons: two are transverse, one is longitudinal, and one is scalar. Although scalar photon states have a negative norm and are worked out with indefinite metric quantization, scalar photon states have positive energy. Then Gupta imposed a condition on the quantum physical states to eliminate the negative norm states. Hence, the physical states are the positive norm states, whereas the negative norm states are auxiliary and do not need physical interpretation.

Gupta's theory was accepted since those negative norm states are just auxiliary and do not appear in the physical space of states. Although Dirac's proposal and interpretation were not accepted, indefinite metric quantization was viewed as legitimate in many approaches. The perturbation theory for gravity requires higher derivatives in the free action, and their presence in the Lagrangian leads to ghost fields, states with negative norm \cite{hahe}. A brief review of physical problems leading to indefinite metric quantization and non-hermitian Hamiltonians was presented in \cite{baga,rami}.

In this regard, {\it i.e.} singularity as an anomaly of the QFT, it has been repeatedly speculated that quantum gravitational field might remove the divergences holding in conventional field theories by providing a natural cut-off associated with the well known fundamental Planck length \cite{dew}. The idea of using a gravitational field for solving the divergence problem of quantum field theory was introduced by Deser in $1957$ \cite{des}. Finally, it has been proven that the singularity of the light cone can only be eliminated by using the quantum metric fluctuation, and some singularities remain. For a review see \cite{for2,for3}.

The negative energy solutions of the field equations are discarded for avoiding negative probability states, but then the symmetrical properties of the field solutions are broken, as was mentioned by Dirac. This fact can be easily seen in the quantization of the massless minimally coupled scalar field in de Sitter space-time. In the procedure of its quantization, the elimination of the negative norm states breaks the de Sitter invariance \cite{dere,gareta}. Since the positive norm states do not form a covariantly closed complete set of solutions of the field equations, the negative norm states for the massless minimally coupled scalar field cannot be eliminated! By comparison, one should notice that the positive norm states in Minkowski space-time form a complete set of solutions, and eliminating the negative norm states does not break the Poincar\'e invariance.

For obtaining a covariant and causal quantization of the massless minimally coupled scalar field in de Sitter space, similar to the Gupta-Bleuler quantization in QED and the ghost fields in quantum non-abelian gauge theory, the negative norm states must be taken into account in the theory and should be considered as auxiliary states. They are not permitted to propagate in the external legs of the Feynman diagrams and can only propagate in the internal line. While the Krein space quantization is applied in a rigorous mathematical way to the free field, the description of an interaction field theory in terms of the Krein space quantization approach remains an open mathematical question \cite{gahure}.

We keep Dirac's and Deser's ideas of renormalizability properties of negative norm states and quantum metric fluctuation, respectively, and combine them with Gupta's idea that the negative norm states are viewed as auxiliary states that are not permitted to propagate in the external legs of the Feynman diagrams. In the previous paper, regardless of the existence of a rigorous mathematical formalism for Krein space quantization in the interaction case, a naturally renormalized QFT was obtained in the one-loop approximation \cite{rota2}. Then we interpreted this result as a new method of regularization in this approximation \cite{fotaza}. In this paper, the ``Krein'' space regularization generalizes to all orders of perturbation theory.

In section \ref{nota}, the notation concerning the singularity problems of the Feynman propagator is given, and then we recall how the singularities of this function are eliminated in the Krein space quantization with quantum metric fluctuations included. Some remarks on Krein space quantization for the interaction case are given in section \ref{kreinqu}. The Krein regularization method is discussed in section \ref{kreinreg}, and the renormalization conditions for the scalar field are reviewed. The effect of this new regularization method on the QFT in curved space-time and quantum linear gravity is discussed. In these cases, there is no need to change the Einstein field equations by the higher derivatives of the metric for the given renormalization of the theory in all orders of perturbation calculation. Finally, concluding remarks and a brief outlook are given in section \ref{conclu}.


\section{Green function} \label{nota}

It is now a well-established fact that the mathematical origin of the divergences in QFT comes from the singular behavior of the Feynman Green function. They may be ascribed to the definition of a point (zero) and infinity on the space-time manifold $\left(\infty\equiv \frac{1}{0}\right)$. However, these concepts cannot be well defined in the quantum field theory. The Feynman propagator for the scalar field is \cite{ful}:
\begin{equation}
\label{fgfh}
\begin{split}
&G_F^p(x,x')=-\ii <0\mid T\phi_p(x)\phi_p(x') \mid 0>=\int \frac{\ud^4
k}{(2\pi)^4}\frac{e^{-\ii k\cdot (x-x')}}{k^2-m^2+\ii \epsilon}
\\ & =-\frac{1}{8\pi}\delta
(\sigma)+\frac{m^2}{8\pi}\theta(\sigma)\frac{J_1
\left(\sqrt{2m^2\sigma}\right)-\ii N_1 \left(\sqrt{2m^2\sigma}\right)}{\sqrt{2m^2
\sigma}} -\frac{\ii m^2}{4\pi^2}\theta(-\sigma)\frac{K_1
\left(\sqrt{-2m^2\sigma}\right)}{\sqrt{-2m^2
\sigma}},
\end{split}
\end{equation}
where $ 2\sigma =\eta_{\mu\nu}(x^{\mu}- x'^{\mu})(x^{\nu} - x'^{\nu})$ is the square of geodesic distance between two points $x$ and $x'$. The divergences of this function come out from its singularity behaviour at short relative distances ($ x-x'\rightarrow 0$), at large relative distances ($ x-x'\rightarrow \infty$) and for light cone propagation ($\sigma= 0$).

The Krein space quantization was used for a massless minimally coupled scalar field in de Sitter space-time for preserving the de Sitter invariance and eliminating the infrared divergence \cite{dere,gareta}. Moreover, we have shown that the quantization in
Krein space also removes all types of divergences in the Green function except the
light cone singularity. In Krein space, the
auxiliary negative norm states (negative frequency solutions that
do not interact with the physical states or the real physical world)
have been utilized. In this case, the decomposition of the field operator into positive and negative norm parts reads as:
\b \phi(x)=\frac{1}{\sqrt{2}}\left[ \phi_p(x)+\phi_n(x)\right]\, ,\e
where the two parts commute with each other and $\phi_p(x)$ is the scalar field as was used in the standard QFT. The ``Feynman'' propagator or the time-ordered product propagator in Krein space quantization is \cite{ta02}:
\b \label{topgf} G_T(x,x')=-\ii <0\mid T\phi(x)\phi(x') \mid 0>=\mathrm{Re}\,G_F^p(x,x').\e
It can be rewritten in terms of Dirac delta, Heaviside and Bessel functions as:
\b \label{fgfk} G_T(x,x')=\int \frac{\ud^4
k}{(2\pi)^4}e^{-\ii k\cdot (x-x')}\mathcal{P}\frac{1}{k^2-m^2}=-\frac{1}{8\pi}\delta
(\sigma) +\frac{m^2}{8\pi}\theta(\sigma)\frac{J_1
\left(\sqrt{2m^2\sigma } \right)}{\sqrt{2m^2 \sigma}},\e
where $\mathcal{P}$ is principal part symbol.

The time-ordered product propagator in Kerin spatial quantization (\ref{topgf}) does not depend on the decomposition of positive and negative components. Since it can be directly constructed from the commutation function of the field operators, which is the same in the Krein space quantization and standard quantization \cite{gareta,rota2,ta02}. However, it is well known that the commutation function is independent of the decomposition into the positive and negative components \cite{ful}. The perturbation theory uses the time-ordered product propagators to calculate the effective action or S-matrix element. Therefore the physical quantities do not depend on a chosen decomposition.

One observes that in Krein space quantization, the two-point function has only the light-cone singularity, and the latter can be removed thanks to the quantum metric fluctuations \cite{for2,for3}. Due to the quantum metric fluctuation, a unique and definite light cone does not exist, and then the Dirac delta singularity disappears. The impact of the quantum metric fluctuation, in the semi-classical approach, on the second part of the Green function (\ref{fgfk}) is negligible. Its impact on the first term, which has a delta-function
singularity on the light cone, however, could not be ignored. Then by considering the QFT in Krein space quantization with included quantum metric fluctuations, it is proved that all singular behaviors of the free scalar Green functions are completely removed \cite{for2,for3,rota2}:
\b \label{fgfk2} \langle G_T(x - x')\rangle =
-\frac{1 }{8\pi} \sqrt{\frac{\pi}{2\langle\sigma_1^2\rangle}}
\exp\left(-\frac{\sigma_0^2}{2\langle\sigma_1^2\rangle} \right)+
\frac{m^2}{8\pi}\theta(\sigma_0)\frac{J_1\left(\sqrt {2m^2
\sigma_0}\right)}{\sqrt {2m^2 \sigma_0}}\, ,\e
where the average is taken over the quantum metric
fluctuations, {\it i.e.} the expectation value of the quantum linear gravity. It is important to note that this Green function is causal and real. In the case of $2\sigma_0 =0$, due to the quantum metric
fluctuation, $h_{\mu\nu}$, $\langle\sigma_1^2\rangle\neq 0$, and we obtain:
\b \label{fgfkm} \langle G_T(0)\rangle = -\frac{1 }{8\pi}
\sqrt{\frac{\pi}{2\langle\sigma_1^2\rangle}} +
\frac{m^2}{8\pi}\frac{1}{2}.\e

A flat background space-time together with a linearised
perturbation $h_{\mu\nu}$ propagating upon it constitute the
basic modality of quantum metric fluctuations {\it i.e.}:
\b \label{linear} g_{\mu\nu}=\eta_{\mu\nu}+h_{\mu\nu},\;\; |h|<|\eta|\, .\e
In the unperturbed space-time, the square of the geodesic distance is defined by $2\sigma_0 =
\eta_{\mu\nu}(x^{\mu}- x'^{\mu})(x^{\nu} - x'^{\nu})$. In a
general curved space-time and in the presence of the perturbation
$h_{\mu\nu}$, the geodesic distance becomes:
\b \label{sigma1} 2\sigma =
g_{\mu\nu}(x^{\mu}- x'^{\mu})(x^{\nu} - x'^{\nu}),\;\; \mbox{and}\;\; \;\sigma = \sigma_0 + \sigma_1 + O(h^2)\,.\e
$\sigma_1$ is the first-order shift in $\sigma$ (an operator in the linear quantum gravity).
It should be noted that $ \langle\sigma_1^2\rangle $ is related to the density of gravitons \cite{for2}. For calculation of $\langle\sigma_1^2\rangle$ see \cite{for3}.

The effect of linear quantum gravity on the scalar Green's function can be calculated by considering the interaction between two fields. For the sake of simplicity, the scalar field with minimal coupling to the gravitational field is considered. In this case, the total classical action reads:
\b \label{grascal} S_c[g,\phi]= \int \sqrt{-g}\; \ud^4x \left(\frac{1}{16\pi G}R+\frac{1}{2} g^{\mu\nu}\partial_\mu \phi \partial_\nu \phi -V(\phi) \right)\,,\e
where $R$ is the scalar curvature and $G$ is the gravitational constant. In the linear approximation of the gravitational field, equation (\ref{linear}), and with the definition $\sqrt{-g}\equiv 1+f(h)$, where $f$ is a differentiable function, we have \cite{car}:
\begin{equation}
\label{linearscal}
\begin{split}
S_c[g,\phi]&\approx \int \left(1+f(h)\right) \ud^4 x \left(\frac{1}{2}\left[ (\partial_\mu h^{\mu\nu})(\partial_\nu h)
-(\partial_\mu h^{\rho\sigma})(\partial_\rho h^\mu_\sigma)\right.\right.
\\
& \left.\left.+\frac{1}{2}\eta^{\mu\nu} (\partial_\mu h^{\rho\sigma})(\partial_\nu h_{\rho\sigma})
-\frac{1}{2}\eta^{\mu\nu}(\partial_\mu h)(\partial_\nu h)
\right] +\frac{1}{2} \left(\eta^{\mu\nu}+h^{\mu\nu}\right)\partial_\mu \phi \partial_\nu \phi-V(\phi) \right)\, .
\end{split}
\end{equation}
For simplicity, we consider only the term $f(h)\eta^{\mu\nu}\partial_\mu \phi \partial_\nu \phi $, which appears in the Lagrangian density (\ref{linearscal}). On the quantum metric fluctuation level it may be written as:
\b \label{conterterm} f(\langle\sigma_1^2\rangle)\eta^{\mu\nu}\partial_\mu \phi \partial_\nu \phi\, .\e
If we assume that $\langle\sigma_1^2\rangle$ is constant, this term can be interpreted as a counterterm for the scalar field. Therefore it modifies the scalar Green's function, and the factor $f\equiv Z_\phi$ plays the role of the wave function renormalization parameter. Similarly, the mass and coupling constant renormalization parameters can be defined. Since $\langle\sigma_1^2\rangle \neq 0$, these terms can eliminate the light-cone singularity (for more details see \cite{dew,des,for2,for3}). 

The other quantum effect can be considered through the quantum effective action, which can be written as:
$$ S_q[g,\phi]= S_c[g,\phi]+\hbar S^{(1)}[g,\phi]+ \cdots \,,$$
where $S^{(1)}$ is the one-loop aproximation.

If we manually impose the condition $\langle\sigma_1^2\rangle=0$, {\it i.e.} removing $\langle\sigma_1^2\rangle$, it reproduces all of the light-cone singularity.
In the limit $\langle\sigma_1^2\rangle \longrightarrow 0$ the delta function is reobtained:
\b \lim_{a \longrightarrow 0} \frac{1}{\sqrt{\pi a^2}}
\exp\left(-\frac{x^2}{a^2} \right)=\delta(x)\, .\e
This means that the fluctuations of the light cone disappear within this limit.
However, we know that at the quantum level and with the quantum metric fluctuation, $\langle\sigma_1^2\rangle$ can not be equal to zero due to the uncertainty principle. More precisely, in QFT, one can not ignore the quantum metric fluctuation effects.


\section{Krein space quantization} \label{kreinqu}

Let us first remember some Krein space quantization features in the interaction case. The Krein space is defined as a direct sum of a Hilbert space and an anti-Hilbert space. The physical states belong to the Hilbert space with a positive norm. In general, the effect of the $S$-matrix with operators acting on the physical subspace of Krein space yields a new state, which is out of the physical space:
$$S|\mbox{Hilbert space}\rangle =|\mbox{Krein space}\rangle\, .$$
Then the inner product of this new state with a physical state result in the projection of this new state onto the Hilbert space as well:
$$S_{if}=\langle\mbox{Hilbert space}|S|\mbox{Hilbert space}\rangle=\langle\mbox{Hilbert space}|\mbox{Krein space}\rangle\, .$$
Let us now analyse this $S$-matrix element $S_{if}$. For the sake of simplicity the $\lambda \phi^4$ theory is discussed. The classical Lagrangian density and $S$-matrix operator are given respectively by:
\b \mathcal{ L}_c(\phi)=\frac{1}{2}\partial^\mu \phi \partial_\mu \phi -\frac{1}{2}m^2 \phi^2 -\frac{\lambda}{4!}\phi^4,\;\;\; S=Te^{\ii \frac{\lambda}{4!}\int \phi^4(x)d^4x}. \e
The effective Lagrangian or ``quantum Lagrangian'' can be established through the loop correction to the classical Lagrangian:
\b \label{qula} \mathcal{ L}_q=\mathcal{ L}_c+ \hbar \mathcal{ L}_1+ \hbar^2 \mathcal{ L}_2+\cdots \, .\e

Since we are interested to the scattering process the connected diagrams are first considered. For the two point function at the order $n$ of the perturbation theory the following integral must be calculated:
$$ \int d ^4x_1\cdots \int d^4x_n \langle q|T \left([\phi(x_1)]^4 \cdots [\phi(x_n)]^4\right) |p\rangle^c= $$
\b \label{tpf} \int d ^4x_1\cdots \int d^4x_n \sum_{i_1\cdots i_n} \langle q| T \phi(x_{i_1})\phi(x_{i_2})|p\rangle G_T(x_{i_3},x_{i_4})\cdots G_T(x_{i_{4n-1}},x_{i_{4n}}) \, ,\e
where the sum is over all possible permitted choices for connected diagrams and $|p\rangle$ and $|q\rangle$ are physical states.

For calculating the time-ordered product, the field operators in this expression must be divided into two types: $1$- those connected to the external line (physical states) and $2$- those connected to the other field operators. The second type produces the time-ordered product Green function (\ref{fgfk}). For the first type, the effect of the negative norm state cancels out the singularity of the positive norm states, then the theory is automatically regularized, and the normal ordering procedure is not needed. One can prove that the effect of the negative norm states on the first part of the equation (\ref{tpf}) cancels out in the one-loop approximation, {\it i.e.} $\langle q|T \phi(x)\phi(x)|p\rangle$, because the unphysical states have negative norm and are orthogonal to the positive norm states. Hence the usual result is obtained without using the normal ordering procedure.

It is important to note that the delta function singularity exists in the second type of products of field operators and that the Krein space quantization does not remove all of the singularity, as was mentioned by Muto et Inoue \cite{muin}. Therefore the effect of the negative norm states on the internal line is simply the replacement of the Feynman Green function (\ref{fgfh}) with the time-ordered product Green function (\ref{fgfk}), and for the external legs, there is the disappearance of the normal ordering procedure.

Although negative norm states appear in Krein space quantization, the negative norm states disappear from the $S$-matrix elements by imposing two conditions. The first condition is the ``reality condition'' in which the negative norm states do not appear in the external legs of the Feynman diagram. This condition guarantees that the negative norm
states only appear in the internal legs and the disconnected
parts of the Feynman diagram. The second condition is that the $S$-matrix elements must be renormalized in the following way:
$$S_{if}'\equiv\mbox{probability amplitude}=\frac{\langle\mbox{physical states}, in|\mbox{physical states}, out\rangle}{\langle 0,in|0,out \rangle}.$$
This condition eliminates the negative norm states in the disconnected parts as well.

In the one-loop approximation, one can prove that the negative norm states disappear, and the usual result is obtained \cite{reta1,reta2,fotaza}. Since the above proof at all orders of the perturbation theory is not yet completed, we change the perspective and use the Krein space quantization, including quantum metric fluctuation, as a new method of quantum field regularisation. In contrast, this property holds at all orders of perturbation theory.


\section{``Krein'' Regularization method} \label{kreinreg}

In QFT, the S-matrix elements or the probability transition amplitude by using the LSZ reduction formula, time evolution operator, and Wick's theorem can be written in terms of a summation and multiplications of the Feynman Green functions (\ref{fgfh}). The singularity appears due to the coincident points and also the multiplication of the Feynman Green functions:
$$ G_F^p(x_i,x_i)\, , \;\; \left[G_F^p(x_i,x_j)\right]^n\,, \;\; \cdots \, .$$
Before presenting our regularization method, we recall that the regularization method in QFT is not unique. However, after imposing the renormalization conditions, a unique result must be obtained for the physical quantities.

Using the above facts, we introduce the ``Krein'' regularization method. The procedure can be completed in two simple steps:
\begin{itemize}
\item[(a)] {\it Replacing the Feynman Green functions (\ref{fgfh}) with the time-ordered product propagator (\ref{fgfk2}).}
\item[(b)] {\it Using the same renormalization conditions as the standard method.}
\end{itemize}
From step (a), it is clear that the theory is entirely finite, and there is no appearance of any singularity since the two-point function is finite and free of any divergences. The theory resulting from this replacement becomes automatically regularized at all orders of the perturbation theory. Therefore by applying this method to the scalar effective action in curved space-time, one can see that there is no need for higher derivative terms for eliminating the divergences, which was previously proved for the one-loop approximation \cite{ta05,entaro}. Hence following this method makes the quantum linear gravity renormalizable.

Using step (a) the $S$-matrix element $S_{if}$ is rewritten as:
$ \langle \alpha, in|\beta, out\rangle_{reg}=f(\alpha, \beta, \langle\sigma_1^2\rangle)$,
where $f$ is a regular function and $\langle\sigma_1^2\rangle $ is density of gravitons. $\langle\sigma_1^2\rangle $ can be considered as a regularization parameter. It is important to note that the $S$-matrix element $S_{if}=f(\alpha, \beta, \langle\sigma_1^2\rangle)$ is finite in all orders of perturbation theory.

The other step (b) guarantees that the physical result does not change. The negative norm states and quantum metric fluctuations for the internal lines eliminate the singularity. Then, by using the renormalization conditions, the effect of the regularization parameter can be absorbed in the redefinitions of the physical parameters, and we obtain the usual result.

For the massive QFT and in the one-loop approximation, one replaces the Feynman Green functions (\ref{fgfh}) with the time-ordered product propagator (\ref{fgfk}). Then, the light cone singularity disappears from the integral representation in the probability amplitude.

Finally, let us recall the renormalization conditions for the effective scalar potential or effective action (\ref{qula}), which is the generating function of the one-particle irreducible diagrams. The effect of the negative norm states and the density of gravitons, which appear in the loop corrections, can be absorbed in the renormalization procedure. The renormalization conditions are imposed through:
\begin{equation}
\label{lambdamu}
\lambda_\mu=-\left.\frac{\delta^4 \mathcal{ L}_q}{\delta \phi^4}\right| _{\phi=\mu}\, , \;\;\; m_\mu^2=-\left.\frac{\delta^2 \mathcal{ L}_q}{\delta \phi^2}\right| _{\phi=\mu} \, , \;\;\; 1=\left. \frac{\delta}{\delta \partial^\nu\phi}\frac{\delta \mathcal{ L}_q}{\delta \partial_\nu\phi}\right| _{\phi=\mu}\, .
\end{equation}
The parameter $\mu$ is the energy scale according to which the mass and the coupling constant are measured.

The effective action in the one-loop approximation for Kerin's regularization approach was previously calculated \cite{reta1}. $\lambda_{\mu}$ is a function of a parameter $\mu$ which is the energy scale of the interaction \cite{reta1}:
$$ \lambda_{\mu}=\lambda
-\frac{\lambda^2}{(8\pi)^2}\left[6\ln\frac{\mu^2}{m^2}+19+12\ln2\right]+O(\lambda^3).$$
For calculating the running coupling constant a scale of energy must be chosen as $\phi=\mu=e^{-t}$. Then the running coupling constant
is defined as $\bar\lambda(t,\lambda)$. The Beta function can be calculated as well \cite{reta1}:
\b \beta=\frac{d\bar\lambda(t,\lambda)}{dt}=\frac{3\lambda^2}{16\pi^2}\, ,\e
which is the same as in the usual results in the one-loop approximation.

The QED effective action has been calculated by using the Krein space quantization, and the quantum metric fluctuations at the one-loop approximation \cite{reta2}. The radiative corrections of QED have also been calculated using the Krein regularization method in this approximation \cite{fotaza}. Furthermore, we have shown that the negative norm states can be used as regularization devices for specific problems \cite{ta02,rota,khnarota,petata,fota}.

One can prove that by using the Green function (\ref{fgfk2}), the effect of the negative norm states disappear up to a one-loop approximation, and it preserves the unitarity in this approximation. The unitary property of the S-matrix is directly linked to the conservation of the probability current in quantum theory and then in observed reality. 

The standard renormalizable QFT is unitary, and it can explain the physical reality. Despite the absence of a mathematical proof of unitarity of Krein QFT to all orders of perturbation theory, it gives us the same results as the standard QFT. Therefore, we have two models, standard renormalizable QFT and Krein QFT, which give us the same physical result for the observed reality. One of them is unitary, and then the other may be unitary! The physical consequence of the unitarity is the conservation of the quantum information, which means that there is no leakage of the information from the physical space of states to the auxiliary nonphysical states. Since the physical results are the same for the two models, there is no loss of information in the Krein space method.

The advantage of Kerin QFT is the renormalizability of the linear quantum gravity equation (\ref{linearscal}), but the standard QFT for this case is not renormalizable. In the Krein space quantization, the effects of negative norm states can be disappeared in physical reality in all orders of perturbation theory, and their effects are only the elimination of the singularities in QFT without changing the physical results.


\section{Conclusions} \label{conclu}

The Krein space quantization for an interaction field has not yet a satisfactory mathematical construction. In this paper, we have changed the perspective, and the Krein space quantization, including quantum metric fluctuation, is considered a new field regularization method. The procedure of this method has been completed in two steps. The first step (a) guarantees that the divergences disappear in the physical quantities at all orders of perturbation and guarantees that we do not need any new terms for the Lagrangian density. This point is essential for QFT in curved space-time and quantum linear gravity. The theory is renormalized in the second step (b) to obtain the expected results. The effective action for the scalar field in curved space-time in this method is regularized in all orders of perturbation theory, and it does not need to change the Einstein field equations in absorbing the singularity of the effective scalar action, contrarily to usual previous methods, where the higher derivatives of metric appear. Our method can be used to calculate physical observables in scenarios where the effect of quantum linear gravity cannot be ignored. In this model, the problem of non-renormalizability of linear quantum gravity can be solved, and the theory can be regularized at all orders of perturbation theory.

\vskip 0.5 cm \noindent {\bf{Acknowledgements}}: The author is grateful to Jean-Pierre Gazeau and Eric Huguet for helpful discussions. We are grateful to the referee for precise and valuable comments. The author would like to thank the Coll\`ege de France and l'Universit\'e de Paris for their financial support and hospitality.

\end{document}